# $NH_3$ adsorption and competition with $H_2O$ on a hydroxylated aluminosilicate surface

Giada Franceschi,*[1] Andrea Conti,[1] Luca Lezuo,[1] Rainer Abart,[2] Florian Mittendorfer,[1] Michael Schmid,[1] and Ulrike Diebold[1]

[1]Institute of Applied Physics, TU Wien, 1040 Vienna, Austria
[2]Department of Lithospheric Research, Universität Wien, 1090 Vienna, Austria

*franceschi@iap.tuwien.ac.at

**Abstract**

The interaction between ammonia ($NH_3$) and (alumino)silicates is of fundamental and applied importance, yet the specifics of $NH_3$ adsorption on silicate surfaces remain largely unexplored, mainly because of experimental challenges related to their electrically insulating nature. An example of this knowledge gap is evident in the context of ice nucleation on silicate dust, wherein the role of $NH_3$ for ice nucleation remains debated. This study explores the fundamentals of the interaction between $NH_3$ and microcline feldspar ($KAlSi_3O_8$), a common aluminosilicate with outstanding ice nucleation abilities. Atomically resolved non-contact atomic force microscopy, X-ray photoelectron spectroscopy, and density functional theory-based calculations elucidate the adsorption geometry of $NH_3$ on the lowest-energy surface of microcline, the (001) facet, and its interplay with surface hydroxyls and molecular water. $NH_3$ and $H_2O$ are found to adsorb molecularly in the same adsorption sites, creating H-bonds with the proximate surface silanol (Si-OH) and aluminol (Al-OH) groups. Despite the closely matched adsorption energies of the two molecules, $NH_3$ readily yields to replacement by $H_2O$, challenging the notion that ice nucleation on microcline proceeds via the creation of an ordered $H_2O$ layer atop pre-adsorbed $NH_3$ molecules.

**Key words**: $NH_3$, AFM, surface acidities, aluminosilicate, hydroxyls

# Introduction

Ammonia (NH$_3$) adsorption on solid surfaces is a matter of intense applied and fundamental research. It is integral to environmental and industrial processes such as NH$_3$ capture for air pollution reduction, NH$_3$-based fertilizer production, hydrogen storage, and environmental chemistry.[1] Moreover, NH$_3$ is abundant in space, and its adsorption on interstellar grains is studied to glean insights into the nitrogen chemistry of the interstellar medium, thought to occur predominantly via gas-grain reactions.[2] Fundamentally, NH$_3$ adsorption processes are regulated by acid-base reactions between different surface sites and NH$_3$. Because of the exquisite sensitivity of NH$_3$ adsorption to the local properties of the surface, adsorption of NH$_3$ is often used to probe the strength and distribution of surface acidic sites.[3] On oxide surfaces, NH$_3$ binding can occur in three main ways: via the lone pair of its N atom to a cation that acts as a Lewis acid; by H-bonding via one of its H atoms to a surface O atom; or by H-bonding via its N atom to an H atom of surface hydroxyl groups (OH). Other feasible ways of ammonia adsorption involve the complete proton transfer from a Brönsted site to create NH$_4^+$, or NH$_3$ dissociation into NH$_2$ (or NH) and OH species.[4]

To ensure a uniform collection of adsorption sites, NH$_3$ adsorption studies are ideally conducted on atomically controlled single crystals prepared under ultra-high vacuum (UHV). To date, these types of studies have been limited to metals[5–8] and various polymorphs of TiO$_2$,[9–13] where NH$_3$ tends to adsorb molecularly on Ti sites via its N lone pair, although dissociation was observed after electron bombardment or on defective surfaces.[9,12] On the other hand, investigations on important materials for NH$_3$ capture, such as metal-organic frameworks (MOFs) and (alumino)silicates, are typically conducted with porous or amorphous samples.[14–16] In such heterogeneous systems, data interpretation becomes challenging due to surface defects, hydroxyls, or ill-defined atom surroundings that can affect local acidities.[12,17] In aluminosilicates, further complexity is given by the details of their Si-Al framework: the specific distribution of Al ions can affect the heat of adsorption of NH$_3$.[16,18] Another important factor is humidity and the water contents of the sample; this can affect NH$_3$ adsorption through cooperative[19] or competitive[20,21] effects between NH$_3$ and H$_2$O, potentially leading to NH$_3$ replacement by H$_2$O even when the adsorption energy of NH$_3$ is stronger than that of H$_2$O.[20] To explore the fundamentals of NH$_3$ adsorption and its interplay with H$_2$O on complex systems such as aluminosilicates, well-defined model systems in the form of atomically characterized single crystals investigated under pristine conditions are helpful.

The importance of understanding the details of NH$_3$ adsorption on aluminosilicates and its interplay with H$_2$O extends to current atmospheric research on ice nucleation (IN) on mineral dust, an important phenomenon affecting the glaciation of clouds with implications for Earth's climate. K-feldspars, particularly the microcline polymorph (KAlSi$_3$O$_8$), are crucial IN agents. Microcline's exceptional IN activity[22–29] is generally rationalized through two main perspectives: one emphasizing surface chemistry, which would promote the formation of H-bonded H$_2$O networks,[30,31] and the other focusing on composition and structural heterogeneities or defects.[32–35] Another important factor is atmospheric chemistry. While travelling in the atmosphere, mineral dust particles are often embedded within aqueous liquid droplets, which can induce preferential dissolution,[36–38] chemical coatings,[39] or net surface charges[40,41] – all of which can impact ice formation. Specifically, it was found that IN activities are substantially enhanced in the presence of dilute $NH_3$- and $NH_4^+$-containing solutions.[42–44] However, the underlying reasons are debated. Current interpretations revolve around the replacement of $K^+$ ions by $NH_3$ or $NH_4^+$ as well as their adsorption on the surface, which may offer oriented H bonds for ice growth,[40,42,44,45] and the nature of $NH_4^+$ ions, which may replace $H_2O$ within the ice network, increasing the configurational entropy of ice and decreasing its overall free energy.[46,47] Experimental investigations on $NH_3$ adsorption on microcline single crystals and its potential interplay with $H_2O$ may enable testing these hypotheses.

This work explores the fundamentals of the interaction of ammonia with microcline feldspar as a prototypical hydroxylated aluminosilicate. Crystalline microcline samples were cleaved in UHV to expose the lowest-energy (001) surface and investigated by atomically resolved non-contact atomic force microscopy (AFM), X-ray photoelectron spectroscopy (XPS), complemented by density functional theory (DFT) calculations. By depositing controlled amounts of $NH_3$ and $H_2O$ molecules at 100 K on microcline (001), it is found that $NH_3$ adsorbs molecularly by creating H bonds with its surface silanol (Si-OH) and aluminol (Al-OH) groups, occupying the same adsorption site as $H_2O$ with an adsorption energy reaching almost that of $H_2O$. When $H_2O$ is deposited onto the surface with pre-adsorbed ammonia, it partially replaces the adsorbed $NH_3$ molecules rather than creating H-bonded networks with the ammonia layer.

## Materials and methods

The experiments were carried out in a UHV setup consisting of two interconnected chambers: A preparation chamber for sample cleaving and XPS (base pressure below $1 \times 10^{-10}$ mbar), and an adjacent chamber for AFM ($1 \times 10^{-11}$ mbar). A natural microcline feldspar from Russia (from Priv.-Doz. Uwe Kolitsch, Natural History Museum Vienna) was characterized ex situ as detailed elsewhere.[31] (001)-oriented grains from the main crystal were mounted onto Omicron-style stainless-steel sample plates and cleaved in the preparation chamber.[31] After cleaving, the samples exhibited strong surface charges, as also observed on other cleaved insulators.[48] To remediate such charge, the samples were irradiated for one minute with X-rays from the XPS setup.

Water (ultrapure deionized water, milliQ®, further purified through three freeze-pump-thaw cycles) and anhydrous ammonia (Linde, 99.999%) were dosed from leak valves while keeping the sample holder on the preparation chamber's manipulator at 100 K. The number of molecules deposited is always expressed with respect to the primitive unit cell of the hydroxylated surface (u.c.; 0.55 $nm^2$). The calibration is based on the amount (partial pressure × time) needed to obtain a coverage of 1 molecule/u.c. as judged by AFM, assuming 100% sticking probability. Warming up the sample to room temperature caused the desorption of all $H_2O$ and $NH_3$ molecules from the hydroxylated surface, as evidenced by XPS and AFM.

XPS was performed with a non-monochromatic dual-anode Mg/Al X-ray source (SPECS XR 50) and a hemispherical analyzer (SPECS Phoibos 100). Spectra were acquired in grazing emission (70° from the surface normal). The intensities and positions of the Al-Kα-excited XPS peaks were evaluated with CasaXPS after subtracting a Shirley-type background. For the display and analysis of the XPS data, an energy correction was applied to all spectra to compensate for the charging of the electrically insulating sample: The Si 2$p$ core-level peak was set to 103.10 eV, as reported in the literature.[49] Fitting procedures and parameters are reported in Section S3. XPS was used to obtain an approximate desorption temperature for $NH_3$. A nominal coverage of 5 $NH_3$ molecules/u.c. was deposited on the UHV-cleaved sample at 100 K. The amount was calibrated based on the dose (pressure × time) needed to observe one $NH_3$ molecule/u.c. in AFM, assuming 100% sticking probability. The sample was warmed to increasingly higher temperatures in steps of 10 K, and XPS spectra (N 1$s$, O 1$s$, K 2$p$, and Si 2$p$ for energy correction) were acquired at each stage. The coverage of the molecular $NH_3$ roughly halved at a temperature between 150 K and 160 K.

The AFM measurements were performed at 4.7 K using a commercial Omicron qPlus LT head and a differential cryogenic amplifier,[50] in constant-height mode. The qPlus AFM sensors ($k$ = 2000−3500 N/m, $f_0 \approx$ 32 kHz, $Q \approx$ 50000) had a separate contact for the tunneling current. Before each measurement, the tips were prepared on an oxygen-exposed Cu(110) single crystal by repeated indentation and voltage pulses. $CuO_x$-terminated tips were prepared on the oxygen-induced reconstruction of Cu(110)[51] to exhibit a frequency shift smaller than −1.5 Hz. Local contact potential difference (LCPD) measurements by the Kelvin parabola method[52] were performed to assess residual fields. Residual surface charges were compensated by applying a bias voltage $V_s$ to the back of the sample plate while keeping the tip potential close to the ground.

DFT calculations were performed with the Vienna Ab-initio Simulation Package (VASP)[53,54] using the r$^2$SCAN-D3 metaGGA exchange-correlation functional.[55] Details about the bulk optimization, unit cell, geometries, cutoff energies, and convergence criteria are specified elsewhere.[31] The AFM images were simulated with the Probe Particle Model,[56,57] which includes Hartree-potential electrostatics and Lennard-Jones potentials as well as the elastic properties of the tip based on the methods described in Refs. [56,57]. $CuO_x$ tips were simulated with the following values of lateral and vertical spring constants and charges: $k_{x,y}$ = 161.9 N/m, $k_z$ = 271.1 N/m, effective tip charge of $-0.05e$. The oscillation amplitude for each simulation always matched the one used in the corresponding experimental image. Since the exact height of the tip is not known, simulated AFM images were calculated for different tip heights; the displayed simulated image is the one fitting the experiment best. Tip-sample distances are always referenced to the most protruding surface atom.

## Results

Previous studies have shown that microcline (001) readily hydroxylates.[31] Even when the sample was cleaved in UHV, the water entrapped in the mineral, and freed during the cleave, is sufficient to create a fully hydroxylated layer that was stable at room temperature.[31] Thus, the starting point of all experiments in this paper was a hydroxylated surface. As shown in Figs. 1a, d, such a layer consists of an ordered array of silanol (Si-OH) and aluminol (Al-OH) groups. When water is dosed below 150 K onto the hydroxylated surface, it adsorbs molecularly at well-defined sites, i.e., in between adjacent silanol and aluminol groups – accepting an H bond from the silanol, and donating one to the aluminol.[31] Figures 1b, e show the DFT-optimized model and the experimental AFM images obtained for a coverage of one $H_2O$ molecule per unit cell (in addition to the dissociated $H_2O$ of the hydroxylated surface in Fig. 1a). In experiments performed with $CuO_x$ tips,[51] the appearance of the water species is very sensitive to the tip-sample distance (Fig. S1a−d). At large distances, $H_2O$ appears with a positive frequency shift (bright); as the tip approaches closer, dark, isolated features suddenly appear. These dark features have a rather sharp boundary and a bright rim, as seen in Fig. 1b.

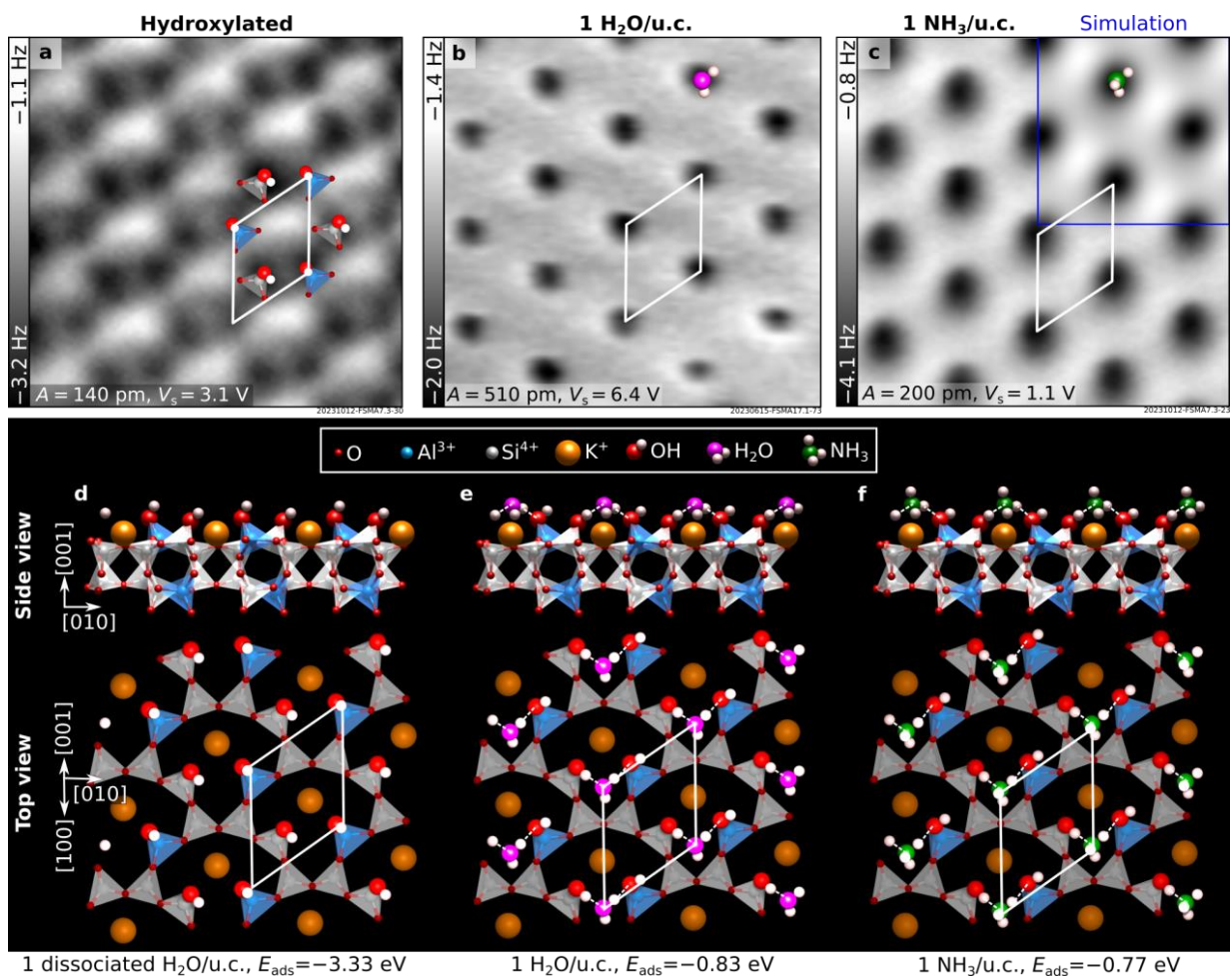

**Figure 1. $H_2O$ and $NH_3$ adsorption on microcline (001).** (a−c) 3.2 × 3.2 nm$^2$ AFM images of the hydroxylated, $H_2O$-, and $NH_3$-exposed microcline (001) surface acquired with CuO$_x$ tips.[51] $H_2O$ and $NH_3$ were deposited at 100 K onto the hydroxylated surface to reach a coverage of one molecule per unit cell (u.c.). (d−f) Corresponding DFT models. SiO$_4$ tetrahedra are gray, AlO$_4$ tetrahedra blue. Panels (d, e) are adapted from Ref. [31]. The AFM simulation obtained from the model in panel (f) is shown in the inset of panel (c); acquired at a tip−sample distance of $d = 5.5$ Å. White rhombi identify primitive unit cells. Calculated adsorption energies ($E_{ads}$) are reported below the corresponding models.

Dosing $NH_3$ below 150 K onto the hydroxylated microcline surface produces a similar structure as the water-exposed one for comparable coverages (Fig. 1c). One dark feature per unit cell is visible, appearing larger and smoother than those caused by $H_2O$ molecules and producing a more gradual frequency change as a function of the tip-sample distance (for a comparison of images of the $NH_3$ and $H_2O$ taken with the same tip, see Fig. S1). Figure 1f presents the lowest-energy structure found for one adsorbed $NH_3$ molecule per unit cell. An intact $NH_3$ occupies the same adsorption site as $H_2O$. The calculated $NH_3$ adsorption energy ($E_{ads}$) is only marginally lower than for $H_2O$ (−0.77 eV vs. −0.83 eV), consistent with the similar desorption temperature of $H_2O$ and $NH_3$ (between 150 K and 160 K, as inferred from XPS – see Methods). Like $H_2O$, $NH_3$ accepts an H bond from a silanol group while donating one to the adjacent aluminol group. One of the two remaining H atoms points away from the surface; the other one lies within the surface and points in the same direction as the free H atom of adsorbed H$_2$O. The corresponding AFM simulation, superimposed on the experimental image of Fig. 1c, reproduces the experimental contrast.

In XPS, the N 1$s$ peak measured on a surface covered by 1 NH$_3$/u.c. sits at 400.4 eV XPS binding energy after correction for charging[31] (see Section S3 for details). The peak intensity increases with ammonia deposition at 100 K while preserving its position and width until the nominal amount of ≈ 3 NH$_3$/u.c., after which the intensity saturates (see Section S3). This suggests that multilayer adsorption is possible only below 100 K, consistent with previous temperature programmed desorption studies on crystalline forsterite (Mg$_2$SiO$_4$)[2] and TiO$_2$,[13] which observed the multilayer NH$_3$ desorption peak around 100 K. Warming up the NH$_3$-covered sample to room temperature causes the intensity to decrease and leads to the eventual disappearance of the N 1$s$ peak without the formation of any shoulders, suggesting that NH$_3$ desorbs from the surface as an intact molecule.

To test whether adsorbed NH$_3$ can serve as a template for an ordered and oriented layer of water ice, H$_2$O was deposited at 100 K on the NH$_3$-saturated surface of Figs. 1c, f. In XPS (Fig. 2a), the H$_2$O component of the O 1$s$ peak (black) increases as a function of the H$_2$O coverage, as expected. However, the increase is significantly smaller than when H$_2$O is deposited directly on the hydroxylated surface (grey), indicating that fewer H$_2$O molecules stick to the surface with pre-adsorbed NH$_3$. (While this could also indicate that 3D clusters are formed, which contribute less to the XPS intensity, the AFM data discard this hypothesis.). At the same time, the N 1$s$ signal (green) decreases. Figures 2b−d show the evolution of selected N 1$s$ and the O 1$s$ peaks (see Section S3 for details about the fitting procedures).

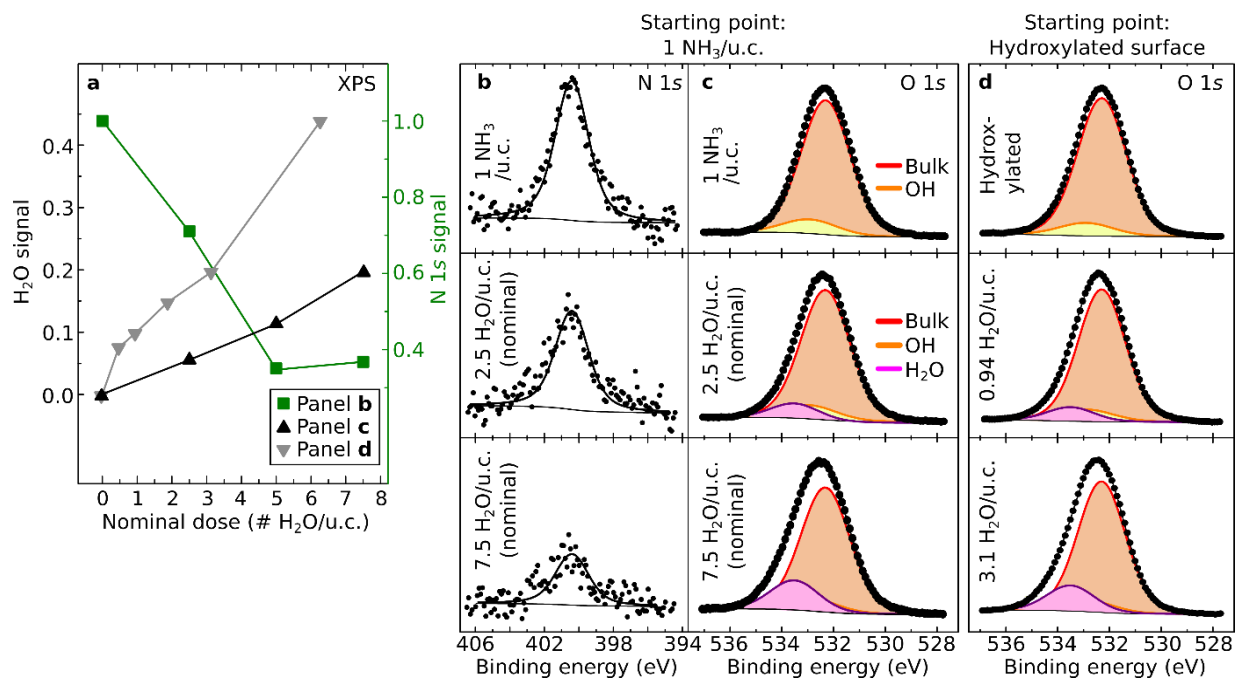

**Figure 2. Substitution of NH$_3$ by H$_2$O measured with XPS.** (a) XPS signals as a function of nominal H$_2$O dose, for adsorption at 100 K onto an NH$_3$-exposed surface (green, squares: N 1$s$, black, upward-pointing triangles: H$_2$O component of O 1$s$), and over a hydroxylated surface without NH$_3$ (grey, downward-pointing triangles: H$_2$O component of O 1$s$). N 1$s$ counts are normalized to the peak area obtained for 1 NH$_3$/u.c.; H$_2$O areas are normalized to the O 1$s$ bulk component of the O 1$s$ peak before any H$_2$O exposure. Error bars on the intensities are comparable to the symbol sizes. The nominal doses do not account for intrinsic errors due to pumping by the chamber walls. Lines connecting symbols are meant to guide the eye. (b−d) Evolution of selected N 1$s$ and O 1$s$ spectra with corresponding fittings. Nominal H$_2$O doses are noted at the sides of the corresponding panels. For all spectra: Al Kα, 1486.61 eV, pass energy 20 eV, 70° grazing emission. Binding energy axes were adjusted to account for charging (see Methods). The O 1$s$ peak was fit as described in Section S3.

The AFM images in Fig. 3 help interpreting the XPS trends. Dosing a nominal amount of 2.5 $H_2O$/u.c. onto the surface with pre-adsorbed $NH_3$ produces the surface shown in Fig. 3a. Most of it appears as light-gray areas with weak modulations of $\Delta f$. When imaged at closer tip-sample distances (Fig. 3d), these areas have the same appearance as the hydroxylated surface with a mixture of $H_2O$ and $NH_3$ molecules (compare with Figs. 1b, c). Similar to the case of single-molecule adsorption, it is evident that also in the mixed phase the $H_2O$ and $NH_3$ molecules occupy the same site of the unit cell. The deposited $H_2O$ has substituted ≈40% of the pre-adsorbed $NH_3$ molecules. In addition to the areas of coexisting $NH_3$ and $H_2O$, a few isolated darker dots are visible in Fig. 3a. Darker contrast indicates a stronger attractive interaction with the AFM tip during the constant-height image acquisition. It is attributed to water species deposited on top of the mixture of $NH_3$ and $H_2O$, thus sticking out further from the surface. In the top part of Fig. 3a, imaged at a closer tip-sample distance in Fig. 3e, a cluster of such darker features is visible. This cluster resembles what is observed by dosing $H_2O$ directly onto the hydroxylated surface without any pre-adsorbed $NH_3$ (Section S2). However, in the absence of pre-adsorbed $NH_3$, the water islands are significantly larger for comparable gas doses, consistent with the higher $H_2O$ signals measured in XPS (Fig. 2a). Additionally, the water islands resulting from deposited water only display internal short-range ordering; such ordering is instead absent in the water islands of Fig. 3.

The water islands grow bigger with increasing water coverages, see Figs. 3b, c, where the nominal gas doses correspond to 5 $H_2O$ and 7.5 $H_2O$/u.c., respectively. In Fig. 3b, the water islands occupy almost the whole surface (the dark dots are not so well resolved as in Fig. 3a because the image is acquired with the tip further away from the surface). A few small, bright patches remain where $H_2O$ and $NH_3$ coexist, two of which are marked. In Fig. 3c, the surface is fully covered by the protruding water islands; several dark spots with different attractive contrast and without any long-range ordering are visible. In XPS (Fig. 2a), the $NH_3$ signal has reached its minimum already at the nominal gas dose of 5 $H_2O$/u.c., indicating that $H_2O$ replaces $NH_3$ only to a certain extent. Thus, $H_2O$ grows atop the mixed layer of $NH_3$ and $H_2O$. Note that it cannot be excluded that a small fraction of $NH_3$ molecules participate to the growth of the water islands mentioned above.

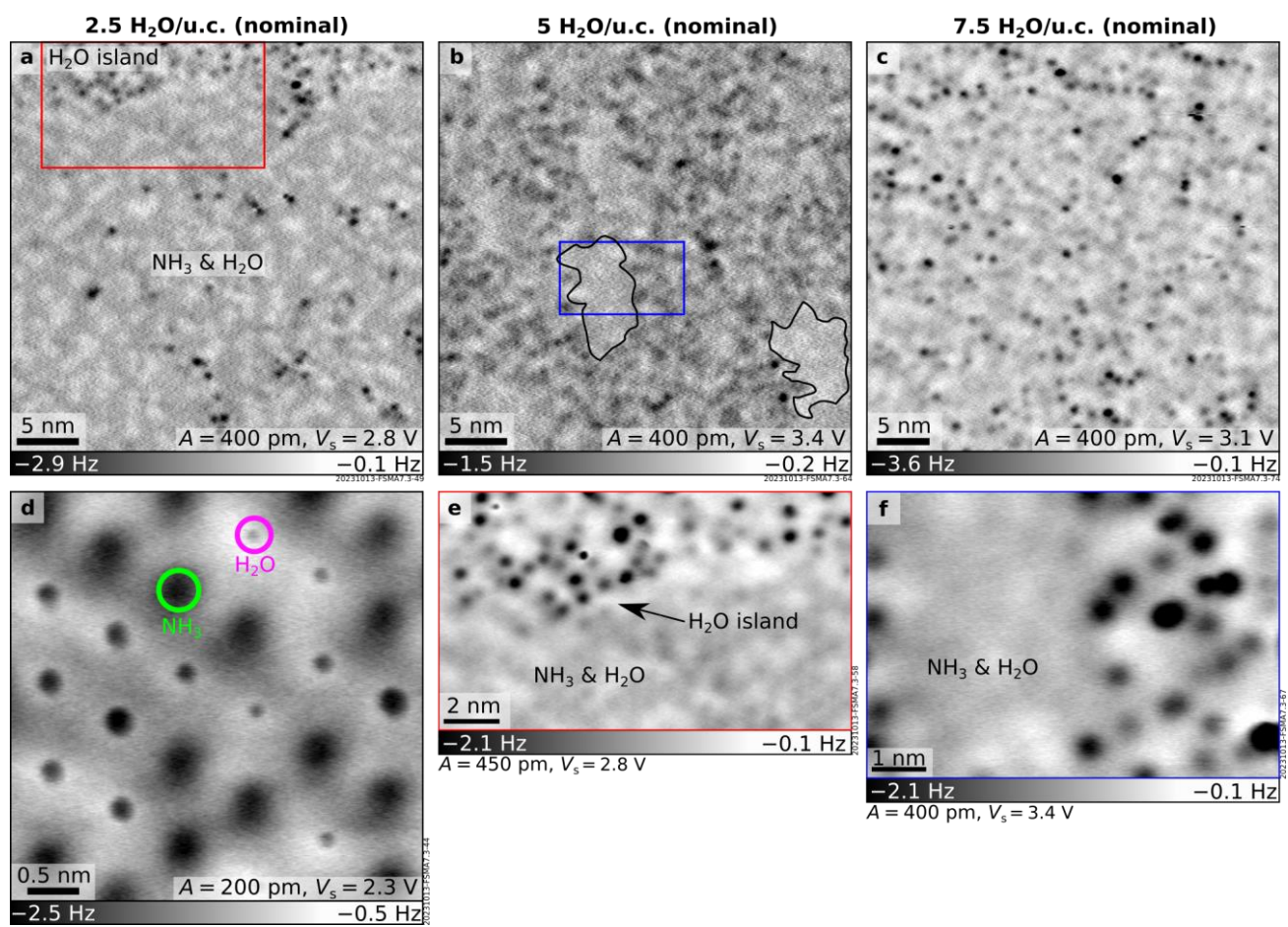

**Figure 3. Substitution of $NH_3$ by $H_2O$, AFM.** (a−f) AFM images of a surface exposed to $NH_3$ at 100 K (gas dose corresponding to 1 molecule per unit cell, as in Figs. 1c, f), followed by $H_2O$ exposure. (a, d, e) Nominal gas dose of 2.5 $H_2O$/u.c.. (a) The majority of the surface appears as a mixture of $H_2O$ and $NH_3$, as seen in more detail in panel (d). The top portion of panel (a), imaged at a closer tip-sample distance in panel (e), evidences the growth of additional $H_2O$. (b, f) Nominal gas dose of 5 $H_2O$/u.c.: The surface is almost fully covered by $H_2O$ islands. Small patches remain in which adsorbed $H_2O$ and $NH_3$ coexist, two of which are marked by dark outlines. The area marked by the blue rectangle is imaged at a closer tip-sample distance in panel (f). (c) Nominal gas dose of 7.5 $H_2O$/u.c.: The surface is fully covered by a disordered $H_2O$ layer.

## Discussion

### Molecular insights into $NH_3$ adsorption.

The (001) surface investigated here is the lowest-energy facet of microcline. Thus, it is likely that this surface is also exposed by mineral dust particles. It serves as a model system to investigate the adsorption mechanisms of gas-phase ammonia and water on hydroxylated aluminosilicate surfaces. The data reveal that $NH_3$ deposited below 150 K adsorbs molecularly via H bonds with the available surface hydroxyls. It is reasonable for molecular adsorption to prevail over the formation of $NH_4^+$ ions or $NH_3$ dissociation upon adsorption. The strong energy gain associated with surface hydroxylation (adsorption energy of $\approx -3.3$ eV/$H_2O$, see Fig. 1b[31]) outweighs the interaction between $NH_3$ and the hydroxyl groups (adsorption energy of $\approx -0.8$ eV, see Fig. 1c), making the transfer of protons from silanols to $NH_3$ to create $NH_4^+$ unlikely. The situation could be different in solution, however, where the ready availability of protons may stabilize $NH_4^+$ ions while maintaining the surface hydroxyls. $NH_3$ dissociation is also not expected because of the lack of available surface O sites that would accept a proton.

The $NH_3$ molecule acts as both a H-bond acceptor and H-bond donor in the adsorption configuration on microcline. According to the general expectation, $NH_3$ should accept H bonds from surface OH groups by forming an H bond between its N atom and the H atom of the OH group.[4,21] Because of the favorable geometry of surface hydroxyls on the hydroxylated microcline surface, $NH_3$ here exhibits amphoteric behavior with respect to H bonds: The $NH_3$ molecule engages with two proximate OH groups by accepting an H-bond from the silanol and donating one to the aluminol. The individual behaviors of silanols (H-bond donating) and aluminols (H-bond accepting) are determined by their coordination chemistry: The larger charge of silicon compared to aluminum (oxidation state +4 vs. +3) strengthens the Si-OH bond relative to the Al-OH bond, making it easier to release H from Si-OH than from Al-OH. These considerations equally apply to the adsorption of $H_2O$, which occupies the same site as $NH_3$ and interacts with proximate OH groups in a similar manner.[31]

The results demonstrate the significance of the atomic environment in determining the acidity of surface hydroxyls, as already predicted by previous computational studies on amorphous silica, where variations in the exact coordination and atomic environment lead to distinct acid behaviors even for the same hydroxyl type.[15,17,58] Extending these reasonings to aluminosilicates, one expects that the relative density and distribution of aluminol and silanol groups (and thus, of Si and Al ions in the Si-Al framework) will affect local acidities. Consistently, zeolites and mesoporous silica evidence changes in heats of adsorption for $NH_3$ and acidity in response to alterations in Al concentrations and distributions.[16,18,59] Therefore, accurate acidity measurements on aluminosilicate surfaces necessitate careful consideration of the distribution and geometric proximity of aluminols and silanols. In this respect, other terminations of microcline besides the (001) surface investigated here, as well as other polymorphs exposing surfaces with different arrangements of silanol and aluminol groups, are expected to produce adsorption configurations of ammonia and water differing from those reported here.

**Interplay between $NH_3$ and $H_2O$ and implications for atmospheric ice nucleation.**
Figures 2 and 3 illustrate how a hydroxylated surface pre-dosed with $NH_3$ changes after exposure to $H_2O$ at low temperatures. $H_2O$ does not simply adsorb atop $NH_3$. Instead it tends to substitute adsorbed $NH_3$, as judged from the coexistence of $H_2O$ and $NH_3$ at equivalent adsorption sites seen in Figs. 3a, d and by the XPS trends of Fig. 2a. For comparable $H_2O$ doses, the $H_2O$ signals are larger for the hydroxylated surface without $NH_3$ compared to the $NH_3$-pre-adsorbed one, suggesting that $H_2O$ sticks less to the latter. Such sticking effects are likely due to the competition of $H_2O$ and $NH_3$ to adsorb on the same site, which promotes the substitution of the adsorbed species over $H_2O$ growth atop $NH_3$. A chemical interaction of $H_2O$ with $NH_3$ is deemed unlikely. If one of the H atoms of $H_2O$ were to interact with $NH_3$ to transform it, e.g., to $NH_4^+$, a shift of $\approx 2$ eV in the N $1s$ peak would be expected,[60] which is not present here. Moreover, an attractive interaction between $H_2O$ and $NH_3$ would favor $NH_3$ to remain on the surface. On the contrary, the $NH_3$ signal decreases with increasing $H_2O$ dose.

This indicates that $NH_3$ and $H_2O$ compete for adsorption sites and speaks against a strong binding between them.

Notably, the substitution occurs despite the minimal energy differences in the adsorption energies of $H_2O$ and $NH_3$ predicted by DFT (−0.83 eV vs. −0.77 eV). Similar effects have been previously observed within MOF cages,[20] where $NH_3$ molecules bound to metal centers can be readily substituted by $H_2O$ even when $NH_3$ has significantly stronger adsorption energies. This behavior was attributed to cooperative effects between the $H_2O$ molecules as well as a strong interaction between $H_2O$ and $NH_3$, which can weaken the bond between $NH_3$ and the metal center while reducing the kinetic barriers to perform the substitution. Similarly, incoming $H_2O$ molecules onto the microcline surface with pre-adsorbed $NH_3$ may interact with the adsorbed $NH_3$ via H-bonding, weakening the bond between $NH_3$ and the surface OH and eventually leading to its detachment. After ≈40% of $NH_3$ has been substituted by $H_2O$, $H_2O$ islands grow atop the mixed $NH_3$−$H_2O$ layer.

The current investigation contributes to ongoing discussions on atmospheric IN on microcline and the role of ammonia in this process. The mechanisms leading to enhanced IN activities of microcline upon immersion in ammonia solutions[42–44] are currently debated. Some studies emphasize the role of the direct adsorption of $NH_3$ or $NH_4^+$ to the surface hydroxyls.[42] In this case, the oriented, extra protons offered by the adsorbed $NH_3$ molecules should favor the formation of ordered H-bonded networks and thus facilitate IN.[40,42,45] This picture is supported by sum frequency generation studies on silica surfaces, indicating that adsorbed $NH_3$ enforces a net orientation of $H_2O$ molecules at the ice-silica interface.[45] The present study shows that, during gas-phase interaction under UHV conditions, $H_2O$ does not orderly bind on top of a layer of adsorbed $NH_3$. Instead, the two species compete for the same adsorption sites, leading to a partial replacement of $NH_3$ by $H_2O$. In the atmosphere, $NH_3$ has a drastically lower concentration than $H_2O$, existing in the parts per billion. Under the oversimplified assumption of gas-phase interactions between microcline and the two substances in the atmosphere, one expects $H_2O$ adsorption to prevail over $NH_3$ because of the concentration difference of the two species and their comparable binding energies. Thus, in this context, the mere adsorption of water over ammonia can be ruled out as a dominant factor for increased IN activities. However, the situation will be different in liquid and thick ice layers. Effects proposed in the literature, such as pH-dependent surface charge, ion exchange ($K^+$ for $NH_4^+$),[44] or preferential dissolution, may dominate. Moreover, some studies suggest that the effect of ammonia solutions is related to the peculiar properties of the $NH_4^+$ ions: $NH_4^+$ is prone to replace $H_2O$ within the ice network while offering an additional proton for H bonding, increasing the configurational entropy of the ice structure and thus lowering its free energy.[46,47]

**Conclusions**

This work sheds light on the molecular-scale interaction of gaseous $NH_3$ with a prototypical aluminosilicate, particularly one prone to hydroxylation upon exposure to water vapor. Based on the phase diagram in Ref. [31], the surface of microcline (001) should be hydroxylated (as in Fig. 1a) under ambient conditions as well. The stability of the hydroxylated surface makes it pertinent for fundamental investigations of environmental processes such as atmospheric ice nucleation. Through atomically resolved AFM imaging and DFT-based calculations, it was found that $NH_3$ adsorbs molecularly on hydroxylated microcline feldspar (001) by H-bonding to its surface aluminol and silanol groups. The geometric proximity and coordination chemistry of the two hydroxyl groups determine the specific adsorption configuration of $NH_3$, which acts as both H-bond donor and H-bond acceptor. The intimate link between the adsorption geometry of $NH_3$ and the type and distribution of surface OH groups evidenced here stresses the importance of determining the distribution of surface Si and Al tetrahedra for reliable acidity measurements of aluminosilicates. When water is introduced onto the surface with pre-adsorbed $NH_3$, it competes for the same adsorption sites, leading to partial replacement and precluding the growth of ordered water networks over adsorbed $NH_3$.

## Conflicts of interest
There are no conflicts to declare.

## Data Availability
The data that support the findings of this study are available from the corresponding author upon reasonable request.

## Acknowledgements
This work was supported by the European Research Council (ERC) under the European Union's Horizon 2020 research and innovation programme (grant agreement No. 883395, Advanced Research Grant 'WatFun'). The computational results presented have been achieved using the Vienna Scientific Cluster (VSC). Priv-Doz. Uwe Kolitsch from the Natural History Museum Vienna is acknowledged for providing the samples used for this work.

## Appendix A. Supplementary Material
Section S1: AFM images at different tip-sample distances. Section S2: Dosing $H_2O$ directly on the hydroxylated surface. Section S3: XPS spectra.

# Supplementary Information for

## NH$_3$ adsorption and competition with H$_2$O on a hydroxylated aluminosilicate surface

Giada Franceschi,*[1] Andrea Conti,[1] Luca Lezuo,[1] Rainer Abart,[2] Florian Mittendorfer,[1] Michael Schmid,[1] Ulrike Diebold[1]

[1]Institute of Applied Physics, TU Wien, 1040 Vienna, Austria

[2]Department of Lithospheric Research, Universität Wien, 1090 Vienna, Austria

*Corresponding author. Email: franceschi@iap.tuwien.ac.at

**This file includes:**

Section S1: AFM images at different tip-sample distances

Section S2: Dosing H$_2$O directly on the hydroxylated surface

Section S3: XPS spectra

Supplementary References

## Section S1: AFM images at different tip-sample distances

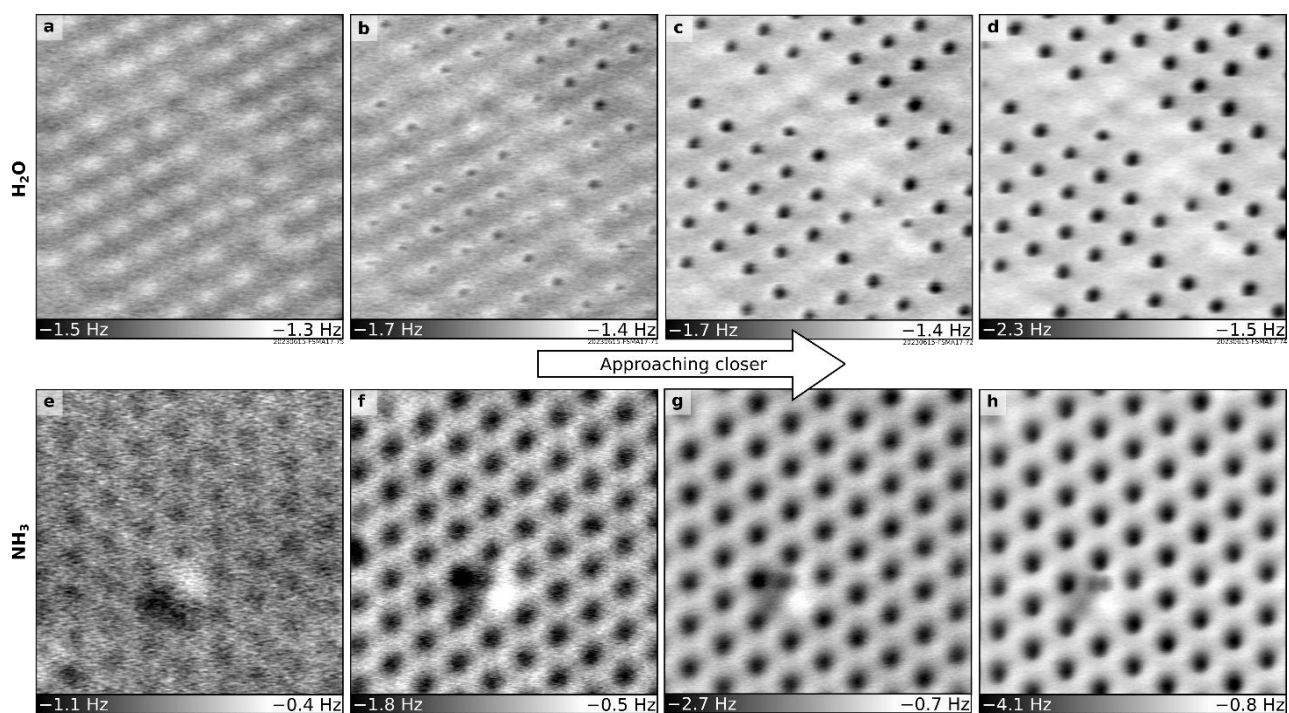

**Figure S1.** AFM images of $H_2O$ (a−d) and $NH_3$ (e−h) dosed at 100 K on a hydroxylated microcline (001) surface at decreasing tip-sample distance.

Figures S1a−d show the appearance of $H_2O$ species with an almost full coverage over a hydroxylated (001) microcline surface. From afar, the species appear bright. At closer tip-sample distances, a small dark dot appears at the center of each bright feature. The dot becomes more pronounced as the tip approaches closer. On the other hand, $NH_3$ species have a much smoother appearance as a function of the tip-sample distance (Figs. S1e−h). Such behavior is also present when the two species coexist on the surface, and allows to distinguish them.

## Section S2: Dosing $H_2O$ directly on the hydroxylated surface

Figure S2 shows the appearance of the hydroxylated microcline surface after a nominal dose of 2.5 $H_2O$/u.c. at 100 K. Most of the surface (brighter contrast) consists of the structure presented in Fig. 1b of the main text, i.e., with 1 $H_2O$/u.c. H-bonded to two proximate silanol and aluminol groups. The top part of the image is decorated instead with 2D clusters of dark (attractive) features, assigned to $H_2O$ islands. Such islands grow bigger with increasing water coverage until they cover the entire surface (without making 3D clusters). Different from the water islands shown in Fig. 3, obtained by dosing $H_2O$ onto a surface pre-dosed with $NH_3$, the water islands shown in Fig. S2 display internal short-range ordering. Note that the areal coverage obtained with a nominal dose of 2.5 $H_2O$/u.c. is significantly larger here compared to the one obtained on the surface pre-dosed with NH$_3$ (Fig. 3a of the main text), consistent with the XPS intensities shown in Fig. 2a in the main text.

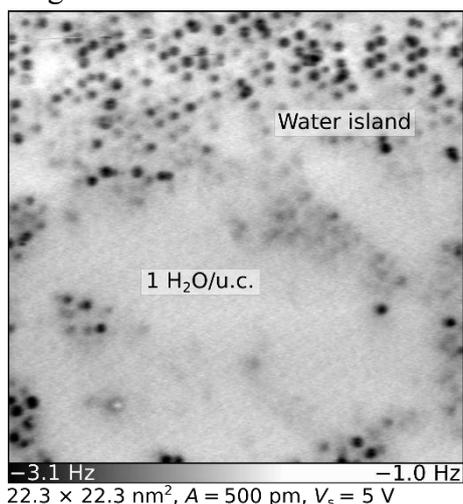

**Figure S2.** AFM image of the hydroxylated microcline (001) surface exposed to a nominal dose of 2.5 H$_2$O/u.c. at 100 K.

## Section S3: XPS spectra

The details of fitting the O 1$s$ peaks are discussed in depth in Ref. 1, and repeated below for convenience. The O 1$s$ peak of the cleaved surface was fit by comparing spectra acquired under normal and grazing emission. The normal-emission spectrum was fit by component 1 alone. Fitting the grazing-emission spectrum required component 2 in addition. Component 2 is assigned to surface OH species that saturate the surface at room temperature. Increasing amounts of H$_2$O at 100 K induced the growth of a third component, which is assigned to molecular H$_2$O. Its position and FWHM were determined from high-dose experiments, which were then constrained to fit the lower doses. For the fits of the low-temperature water experiments, the intensity ratio of components 1 and 2 was constrained to the value found on the cleaved surface. This is based on the assumption that molecular H$_2$O grows onto the fully hydroxylated surface. Table S1 summarizes the relevant fitting parameters.

**Table S1. Details about the XPS fitting components.** The shape (LA=asymmetric Lorentzian), full-width half maximum, and position were constrained for all peaks.

|  | Identifier | Shape | FWHM | Position (eV) | Area |
| --- | --- | --- | --- | --- | --- |
| O 1$s$ 1 | O 1$s$ cleaved | LA(1.53,243) | 2.24 | 532.30 | Free |
| O 1$s$ 2 | OH cleaved | LA(1.53,243) | 2.5 | (O 1$s$ 1) + 0.60 | (Area O 1$s$ 1) × 0.117 (for molecular H$_2$O dosing) |
| O 1$s$ 3 | H$_2$O | LA(1.43,243) | 2.2 | (O 1$s$ 1) + 1.20 | Free |
| N 1$s$ | N 1$s$ | LA(1,253) | 2.3 | 400.4 | Free |

Figure S3 shows the evolution of the N 1$s$ XPS peak upon dosing NH$_3$ on the hydroxylated microcline surface at 100 K. The peak intensity (Fig. S3a) steadily increases up to a nominal dose of ≈ 3 NH$_3$ molecules per unit cell, after which it saturates (Fig. S3b).

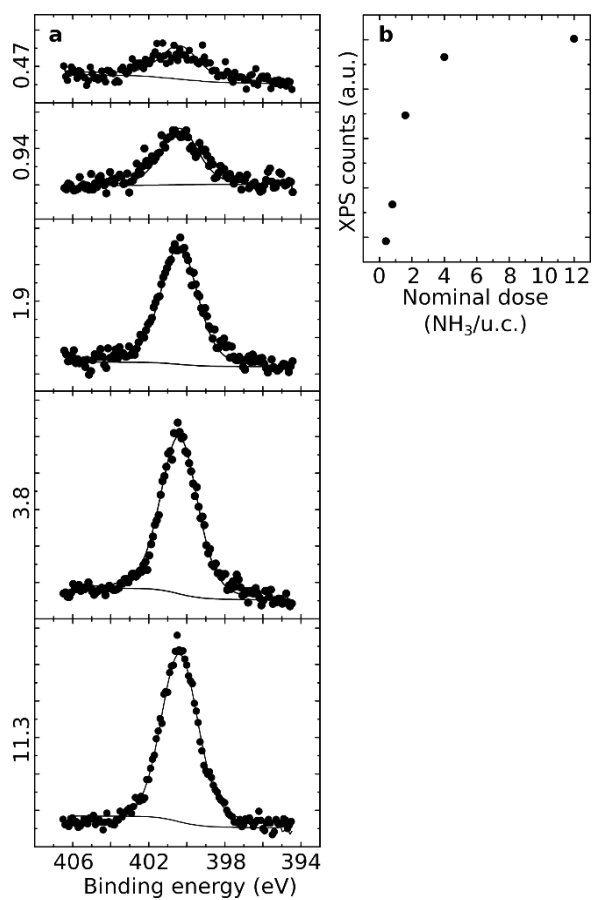

**Figure S3. Evolution of the N 1$s$ peak as a function of NH$_3$ dose at 100 K.** (a) Experimental data (circles) and fits (solid lines) of the N 1$s$ peak as a function of the nominal NH$_3$ dose, expressed as number of NH$_3$ molecules per unit cell (numbers at the left). (b) Plot of the corresponding intensities. For all spectra: Al K$\alpha$ radiation, 1486.61 eV, pass energy 20 eV, 70° grazing emission. The binding energy axis was adjusted to account for charging (see Methods).

## Supplementary References